
\magnification=\magstep0
\hsize=13.5 cm               
\vsize=19.0 cm               
\baselineskip=12 pt plus 1 pt minus 1 pt  
\parindent=0.5 cm  
\hoffset=1.3 cm      
\voffset=2.5 cm      
\font\twelvebf=cmbx10 at 12truept 
\font\twelverm=cmr10 at 12truept 
\overfullrule=0pt
%
\newtoks\leftheadline \leftheadline={\hfill {\eightit Authors' name}
\hfill}
\newtoks\rightheadline \rightheadline={\hfill {\eightit the running title}
 \hfill}
\newtoks\firstheadline \firstheadline={
\hfill}
\def\makeheadline{\vbox to 0pt{\vskip -22.5pt
\line{\vbox to 8.5 pt{}\ifnum\pageno=1\the\firstheadline\else%
\ifodd\pageno\the\rightheadline\else%
\the\leftheadline\fi\fi}\vss}\nointerlineskip}
%
\font\eightrm=cmr8  \font\eighti=cmmi8  \font\eightsy=cmsy8
\font\eightbf=cmbx8 \font\eighttt=cmtt8 \font\eightit=cmti8
\font\eightsl=cmsl8
\font\sixrm=cmr6    \font\sixi=cmmi6    \font\sixsy=cmsy6
\font\sixbf=cmbx6
%
\def\eightpoint{\def\rm{\fam0\eightrm}
\textfont0=\eightrm \scriptfont0=\sixrm \scriptscriptfont0=\fiverm
\textfont1=\eighti  \scriptfont1=\sixi  \scriptscriptfont1=\fivei
\textfont2=\eightsy \scriptfont2=\sixsy \scriptscriptfont2=\fivesy
\textfont3=\tenex   \scriptfont3=\tenex \scriptscriptfont3=\tenex
\textfont\itfam=\eightit  \def\it{\fam\itfam\eightit}%
\textfont\slfam=\eightsl  \def\sl{\fam\slfam\eightsl}%
\textfont\ttfam=\eighttt  \def\tt{\fam\ttfam\eighttt}%
\textfont\bffam=\eightbf  \scriptfont\bffam=\sixbf
\scriptscriptfont\bffam=\fivebf \def\bf{\fam\bffam\eightbf}%
\normalbaselineskip=10pt plus 0.1 pt minus 0.1 pt
\normalbaselines
\abovedisplayskip=10pt plus 2.4pt minus 7pt
\belowdisplayskip=10pt plus 2.4pt minus 7pt
\belowdisplayshortskip=5.6pt plus 2.4pt minus 3.2pt \rm}
%
%
\def\leftdisplay#1\eqno#2$${\line{\indent\indent\indent%
$\displaystyle{#1}$\hfil #2}$$}
\everydisplay{\leftdisplay}
%
\def\frac#1#2{{#1\over#2}}


%
%
\def\pmb#1{\setbox0=\hbox{$#1$}\kern-0.015em\copy0\kern-\wd0%
\kern0.03em\copy0\kern-\wd0\kern-0.015em\raise0.03em\box0}
%
\def\teff{T$_{\rm eff}$}
\def\logg{$\log g$}
\def\met{[M/H]}

\def\xv{{\bf x}}

\def\sig68{$\sigma_{68}$}
\def\sigrms{$\sigma_{RMS}$}

\def\mae{$\epsilon$}

%
%
%
\pageno=1
\vglue -28 pt
\leftline{\eightrm To appear in  {\eightit Automated Data Analysis in Astronomy},
	  R.\ Gupta, H.P.\ Singh, C.A.L.\ Bailer-Jones (eds.),} 
\leftline{\eightrm Narosa Publishing House, New Delhi, India, 2001}
\vglue 68 pt  
%
%
\leftline{\twelvebf Automated stellar classification for large surveys:}
\vskip 4pt
\leftline{\twelvebf a review of methods and results}
%
\smallskip
\vskip 46 pt  
\leftline{\twelverm Coryn A.L.\ Bailer-Jones} 
\vskip 4 pt
\leftline{\eightit Max-Planck-Institut f\"ur Astronomie, K\"onigstuhl 17, 69117 
Heidelberg, Germany}
\leftline{\eightit email: calj@mpia-hd.mpg.de}
%
%
\vskip 20 pt 
%
%
\leftheadline={\hfill {\eightit Bailer-Jones\hfill}}
\rightheadline={\hfill {\eightit A review of automated stellar classification
}  \hfill}

%
{\parindent=0cm\leftskip=1.5 cm

{\bf Abstract.}
Current and future large astronomical surveys will yield
multiparameter databases on millions or even billions of objects. The
scientific exploitation of these will require powerful, robust, and
automated classification tools tailored to the specific survey.
Partly motivated by this, the past five to ten years has seen a
significant increase in the amount of work focused on automated
classification and its application to astronomical data. In this
article, I review this work and assess the current status of automated
stellar classification, with particular regard to its potential
application to large astronomical surveys. I examine both the
strengths and weaknesses of the various techniques and how they have
been applied to different classification and parametrization problems.
I finish with a brief look at the developments still required in order
to apply a stellar classifier to a large survey.

}                                 

\vskip 20 pt
\centerline{\bf 1.\ Introduction}
\bigskip
\noindent
At its most general level, the objective of classification is to
identify similarities and differences between objects, with the goal
of efficiently reducing the number of types of objects one has to deal
with. Ideally, the classes so produced are motivated by a scientific
understanding of the objects. How we group the objects into classes
depends on many things, including how many classes we want, what
measurement features we have available and what procedure we use to
discriminate between the objects. Of course, at some level of detail,
all objects are unique, but the point is that different aspects of
this uniqueness will be irrelevant in different classification
contexts.

In the case of stellar astrophysics, a classification system
has emerged over the 140 years since Lewis Rutherfurd first divided
stars into three groups based on their  low resolution optical
spectra. While still closely tied to the optical spectra, stellar
classification now reflects underlying physical properties, in
particular the effective temperature, \teff, surface gravity, \logg,
and metallicity, \met. Bright main sequence and giant branch stars can
be represented well by the MK system (Morgan, Keenan \& Kellman 1943),
a two-parameter system in which the spectral type (SpT) is closely
related to \teff, and the luminosity class (LC) is related to
\logg.  This is a fairly coarse system, however, and many ``peculiar'' types
of objects appear as exceptions which cannot be usefully described by
these two parameters alone.

Physical stellar parameters (mass, age, radius, temperature etc.)\ show
continuous distributions, so it is often more appropriate to
parametrize spectra on continuous parameter scales rather than
classify them into discrete classes. For example, the MK spectral
types were originally designated as the classes which were discernible
at a certain wavelength resolution, but we now know them to be
somewhat arbitrary divisions on what is really an underlying
continuous temperature scale.  While I draw this distinction between
classification and parametrization, I will nonetheless refer to the
collective task of determining quantities from spectra as
``classification'' for brevity.

In this article, I give an overview of automated stellar
classification as a tool for large surveys. ``Large'' here means in
excess of one million objects. I start in section 2 with an overview
of the goals of classification in this context, before going on to
review the main classification methods (section 3) and give a critical
comparison of these methods (section 4). I then give an overview of
the literature which illustrates the application of these methods
(section 5).  After summarizing the current state of classification
performance (section 6), I conclude with my view of how automated
classification needs to develop to tackle the challenges posed by
large surveys.

\vskip 12 pt
\centerline{\bf 2.\ The goals of automated stellar classification}
\bigskip
\noindent
Recent technological developments in electronic detectors, but also in
powerful computers and software, have meant that astronomical surveys
of over 10$^9$ objects (e.g.\ the entire sky down to 20th magnitude)
are now being designed and implemented.  The full exploitation of the
data from such surveys clearly requires automated methods.  This is
particularly the case for multiband surveys which will observe objects
in many filters (or even with a low resolution spectrograph) and thus
produce many measurements per object which cannot be summarized in one
or two colour--magnitude diagrams.

A variety of large surveys are underway or being proposed, including
multicolour surveys (e.g.\ 2MASS, DENIS and SDSS), the large-area
synoptic survey and ambitious parallax missions such as FAME, DIVA and
GAIA (see Clowes, Adamson \& Bromage 2001 for an overview). Many of
these are large area, magnitude-limited surveys, so even if their
primary goal is not Galactic astrophysics, they will nonetheless
observe large numbers of stars.

Large surveys are concerned with two things. The first is finding
unusual objects. These will be discovered by virtue of being isolated
from most objects in some parameter space (provided the measurement or
classification system provides this separation). Once detected, these
unusual objects must always be analysed individually, no matter what
classification system is used.  The second goal of a survey is to do
statistics with large numbers of objects, and for this purpose an
automated classification system is required which can extract
information relevant to the astrophysical goals.  I therefore define
the goal of automated stellar classification as {\it the reliable and
precise determination of the intrinsic properties of a wide range of
stars from their spectral energy distributions by predominantly
non-interactive means}. That the classification should be precise
(have low random errors) is self-evident. However, we must be aware
that there are intrinsic limits to how precisely we can determine any
idealized parameter, and that the ``cost'' of achieving ever higher
precision increases rapidly (e.g.\ in terms of collecting area or
integration time). ``Reliable'' refers to low systematic errors,
implying a technique which will not give wild answers when it is
unsure.  For example, a stellar classifier should say that it does not
know what a quasar is rather than just classify it as, say, a G2 star.

The fundamental intrinsic properties of a star are its mass, age and
abundances. Related to these are its radius, effective temperature and
surface gravity. There are also a number of ``secondary parameters''
for characterizing rotation, chromospheric and coronal activity,
microturbulence etc. Many of these can only be measured (or rather
inferred) in specific spectral ranges or at certain resolutions and
signal-to-noise ratios (SNRs). Thus the design of an observational
system depends on which parameters are to be measured.  (Many of these
quantities are not directly observable, so physical modelling plays an
important role, for example in the determination of the mass of an
isolated star.) There are other parameters which are extrinsic to the
star, including its distance, kinematics, interstellar extinction and
companionship.  These may be of immediate interest, or may interfere
with the determination of other parameters.

My above statement argues that an automated classification system
should operate on a wide class of objects. Large surveys will not be
of preselected objects. Thus a classification system which only
operates on a small subset of stellar types will require that so much
effort be put into a reliable preselection system that this system
will already have done a significant part of the classification. Of
course, one can imagine a hierarchical system with progressive stages
of class detail, but, considered as a whole, this system still has to
be applicable to a wide range of stellar types. Much of the early work
on automated classification focused on a limited set of spectral
types, and while this work was important for demonstrating the
techniques, the classification systems produced cannot be directly
implemented for larger surveys.

The final aspect of my defined goal refers to ``predominantly
non-interactive''. That classification for over one million objects
based on tens of measurements per object cannot be done by hand is
obvious, yet we must not delude ourselves into thinking that an
automated system will never fail. There will always be cases with
which the system will have problems, and the skill is to produce a
system which will fail gracefully and inform us when difficulties
arise (or better, quantify its own uncertainties).  This is a
challenge because even a system with a 99\% correct classification
rate will still make ten million errors on a data set of 10$^9$
objects.

This review only looks at existing classification systems, and hence
at {\it supervised} classification methods. {\it Unsupervised}
methods -- which find ``natural'' groupings in a dataset without
reference to externally specified classes -- do not appear to be as
useful for fulfilling the goal described. This is because the new
classes they discover (or rather invent) would still have to be
calibrated and understood in terms of stellar astrophysics. It seems
more sensible to me to use a supervised technique which will classify
objects directly in terms of our physically motivated classes or
parameters. Unsupervised techniques may, however, have a roll to play
as a preprocessor in discriminating known objects from unknown ones.

In the rest of this paper I will frequently refer to the measured
spectrum as the measured feature vector which is used as the basis for
classification. However, this can equally well refer to a set of
non-contiguous flux measurements obtained through a set of filters,
and possible even contain other relevant measurements.

\vskip 12 pt
\centerline{\bf 3.\ Classification methods}
\bigskip
\noindent
Almost all of the recent work on automated stellar classification has
used one of four techniques: principal component analysis (PCA);
neural networks (NN); minimum distance methods (MDM); Gaussian
probabilistic models (GPM). PCA forms a set of linearly independent
basis vectors with which to describe the data, and can be useful as a
classification system by using only the most significant few
components. PCA is described in the article by Singh, Bailer-Jones
\& Gupta in these proceedings.  The neural networks used in automated stellar
classification have almost exclusively been feedforward networks.
These are networks which can be trained to give a mapping between the
stellar spectral domain and the classification parameter domain. See
the article by Bailer-Jones, Gupta \& Singh in these proceedings for
an introduction to these models. The other two methods are now
described in more detail.

\vskip 12 pt
\centerline{\bf 3.1.\ Minimum distance methods (MDM)}
\bigskip
\noindent
Metric distance minimization (also called a minimum distance method)
classifies objects by minimizing some distance metric between the
object to classify and each member of a set of templates. The object
is assigned the class of the template which gives the smallest
distance (closest match). If ${\bf X} = (x_1, x_2, \ldots, x_i,
\ldots, x_N)$ is the feature vector (spectrum) to classify, and ${\bf S}_c =
(s_1^{\{c\}}, s_2^{\{c\}}, \ldots, s_i^{\{c\}}, \ldots, s_N^{\{c\}})$
is a template $c$, we evaluate

$$ D_c = \frac{1}{N} \left[ \sum_{i=1}^{i=N} w_i^{\{c\}} |x_i -
s_i^{\{c\}}|^p \right ]^{1/p} \eqno(1) $$

\noindent 
where $w_i^{\{c\}}$ is a weight assigned to flux element $i$ of that
class $c$.  ${\bf X}$ is assigned to class $c$ for which $D_c$ is
minimum.  The value of $p$ determines the type of distance: typically
$p=2$ is used, which is the normal Euclidean distance metric.  With
this approach, our highest class resolution is set by the grid of
templates, i.e.\ we make a discrete classification equal to one of the
templates. We can improve this by interpolating between the lowest few
values of $D_c$ and by making an inter-class assignment.  Generally we
must determine the weights according to the relative importance of
spectral features for determining various classes. Taking
$w_i^{\{c\}}=1$ for all $i$ and $c$ is generally a poor choice for
stellar classification, as it will attach most significance to the
strongest lines, which are often not the most relevant for
classification.

MDM can be considered as a specialization or generalization of a
number of other methods. It is very similar to the ``$k$ nearest
neighbours'' (or $k$-nn) method, in which an object is assigned a
class based on the classes of its $k$ nearest neighbours in the
feature space. (The feature space is the $N$ dimensional space
containing the measured feature vector.) MDM
without class interpolation is the same as $k$-nn with $k=1$.  If
inter-class assignments are meaningful, then the assigned class could
be a weighted average of the classes of the $k$ nearest neighbours,
with the weights set inversely proportional to the distance to these
neighbours.  In MDM, the class templates are best formed from an
average of a number of examples of a given class, while in the $k$-nn
method this averaging is done specifically for each new object we want
to classify.

MDM is the same as $\chi^2$ minimization when $p=2$ and $w_i^{\{c\}} =
\sigma_i^{-2}$ (for all $c$), where $\sigma_i$ is the error in $x_i$
(which includes the photon noise, calibration errors etc.)\ and the
templates are assumed to be noise free. However, this is not the most
useful weighting for stellar classification, as it is unrelated to the
relative importance of the spectral features in distinguishing between
classes.  MDM is also similar to cross-correlation of ${\bf X}$ on the
templates, differing in the treatment of boundaries (i.e.\ how we
treat the correlation sum when the spectra do not overlap fully).

\vskip 12 pt
\centerline{\bf 3.2.\ Gaussian probabilistic models (GPM)}
\bigskip
\noindent
A different approach is to consider the classification problem in 
terms of probabilities. Let $p(\xv|c)$ be the probability
that a member of class $c$ has feature vector $\xv$.
From Bayes' Theorem, the probability that an object with
a measured feature vector $\xv$\ is a member of class $c$ is

$$ p(c|\xv) \propto p(\xv|c)p(c) \ \ . \eqno(2) $$

\noindent
We can then make the simplifying assumption that
$p(\xv|c)$ is a multivariate Gaussian distribution with mean ${\bf
\mu}$ and covariance matrix ${\bf \Sigma}$.  These can be determined
from a set of preclassified data by various methods (a process we can
refer to as training).  $p(c)$ is the prior probability that an object
is a member of class $c$. We may not have any idea what this is, so
may want to assign an uninformative prior, i.e.\ $p(c)$
constant. However, we may already have some knowledge that this object
is more likely to be a particular type of star (e.g.\ that it is
likely to be low metallicity based on its kinematics). All
classification methods have such a prior, but not all allow us to
specify it (easily). For example, MDM implicitly assumes that $p(c)$ is
constant.

This direct probabilistic approach to stellar classification has been
explored relatively little in the context of stellar classification,
yet a large literature exists on this class of model.  For an example
of an unsupervised approach for the classification of IRAS LRS
sources, see Goebel et al.\ (1989).  One feature of these models which
may often be useful when data are missing (which is inevitable for
large surveys) is the ability to {\it marginalize} over unmeasured
features, $\xv_u$, and classify only on the basis of the measured
features, $\xv_m$:

$$ p(\xv_m|c) = \int p(\xv_m, \xv_u | c) d\xv_u = \int
p(\xv_m|\xv_u,c) p(\xv_u|c) d\xv_u \ \ . \eqno(3) $$

\noindent Gaussian distributions are convenient here, because the marginal
distribution of a multivariate Gaussian is another multivariate
Gaussian.

\vskip 12 pt
\centerline{\bf 4.\ A comparison of the classification methods}
\bigskip
\noindent
\vskip 12 pt
\centerline{\bf 4.1.\ Training}
\bigskip
\noindent
All of the four methods described in the previous section are in some
sense supervised, i.e.\ they assign classifications based on some
preclassified data.  NN and GPM must be explicitly trained, and
encapsulate this training information in their internal parameters
(the network weights and mean/covariance matrix respectively).  With
PCA, the training can be considered as the matrix diagonalization
required to determine the principal components (PCs).  PCA as it
stands is not really a supervised classification method, as the
formation of the PCs is independent of the class assignments. However,
applications in the literature of PCA as a classifier generally then
assumes that the classification is a simple function of the first few
most significant PCs (the function being solved by simple
regression/interpolation techniques).

\vskip 12 pt
\centerline{\bf 4.2.\ Data requirements and speed}
\bigskip
\noindent
MDM does not have to be trained when weighting is not used, as the
training data themselves are retained when making classifications.
This introduces potentially serious problems when we wish to apply the
method to multiparameter problems.  If no interpolation between
classes is used, then the training data should be ``dense'' in each
parameter so that the method can recognise the effects of all
parameters on the feature vector. Thus if we have templates for 30
spectral types, but then want to extend classification to 10
luminosity classes, we need another $30\times(10-1)$ training spectra
so that every class combination is represented: the training data
requirement increases exponentially with the number of
parameters. Moreover, every time we want to classify a new object, we
must evaluate $D_c$ for every template. This will require a
significant amount of computer time if several parameters are
involved, and may be prohibitive for a large survey.  We would
probably want to be able to classify everything in the survey in about
one day, because repeated classification is inevitable as the data or
models are improved. This means each object must be classified within
0.1 milliseconds.  Let us assume that the feature vector has 50
dimensions and that we wish to classify in five parameters. Of these,
one (probably temperature) is represented by 30 different classes, and
the other four by only ten classes each.  The MDM training set would
have to consist of $30\times 10^4$ training spectra, and each new
classification would require 15 million calculations of the form
$(x-s)^2$ to be completed in 0.1 milliseconds.  This is five times
faster than is possible with a CRAY T90 supercomputer.

The logical solution to this problem is to reduce the amount of data
by interpolation. This is possible with MDM through an interpolation
in the class space, as mentioned in section 3.1. This, however,
requires assumptions about the continuity and smoothness between
neighbouring classes which may not be valid. Interpolation in the
feature space, on the other hand, is more robust, as the features are
typically photon fluxes. An interpolation in the feature space is
precisely what neural networks do.  If a 50:10:10:5 network could
solve the above problem, it would require only about 700 products and
50 exponentials to be calculated per object.  Furthermore, neural
networks naturally lend themselves to being programmed in parallel, or
even built into hardware in parallel, in which case the number of
operations is reduced to about 70 products and three
exponentials. This is easily achievable in 0.1 milliseconds.
Additionally, the exponentiation (which is slow) could be done via a
look-up table. Of course, neural networks must be trained. But as they
only have to be trained once per classification run, and are trained
on a much smaller amount of data than the entire survey, the training
could be completed within one day. For example, a 560:5:10:3 network
trained on a few thousand spectra for a similar problem to the example
above took less than one day to train on a Sun Enterprise workstation,
without any parallelization or look-up tables. GPM and PCA have
similar speed characteristics as neural networks, i.e.\ they are
relatively slow to train yet quick to apply.  Note that any
interpolation method (indeed any method with free parameters) requires
that sufficient data be used for the proper determination of these
parameters (see section 5.5 of Bailer-Jones et al.\ in these
proceedings). As training data are generally ``expensive'' to acquire,
it will be desirable to keep the model as simple as possible (although
no simpler).

\vskip 12 pt
\centerline{\bf 4.3.\ Knowledge encapsulation}
\bigskip
\noindent
A neural network encapsulates the information about the different
classes in a single set of weights, yet all of these weights are
involved in the classification of objects of any one class. This means
that the presence of, say, B stars in the training data may affect the
classification of, say, M stars.  In an extreme case this is certainly
true: if we train a network on 1000 B stars and only one M star, we
would expect the network to learn very little about M stars and hence
do poorly at classifying them. Thus we must pay attention to the
relative frequencies of objects in the training data.  For more evenly
distributed training data we are faced with the conceptual problem of
whether the nature of B stars should affect the ability of the network
to classify M stars.  In at least one case, a network trained on a
large range of spectral types showed some internal specialization,
i.e.\ certain hidden nodes specialized to recognise certain ranges of
spectral types: see section 7 of Bailer-Jones et al.\ (these
proceedings).

In MDM, the classification knowledge remains in the templates, all of
which must be retained to make classifications. However, some
additional information may be provided with an appropriate weight
vector.

\vskip 12 pt
\centerline{\bf 4.4.\ Model complexity}
\bigskip
\noindent
PCA is a linear transformation of the data, so any linear
classification model which uses a fit to the first few PCs is likely
to be too simplistic for multiparameter classification. Neural
networks, on the other hand, can be made arbitrarily nonlinear (in
principle), and have a convenient means for investigating degrees of
complexity and nonlinearity (through the use of different numbers of
hidden nodes and with regularization techniques). Both GPM and MDM are
nonlinear, by which I mean that the class is a nonlinear function of
the feature vector. With MDM, the only apparent flexibility of the
model with regard to complexity control (other than by increasing the
amount of data) is the degree of sophistication in the class
interpolation scheme. GPM models are limited by the fact that they
assume Gaussian distributions, although this could be relaxed at the
expense of computational effort.

Not all objects within a given class are identical, so it is necessary
for a classification technique to realise that there are {\it
intra}class differences as well {\it inter}class ones. NN and GPM
methods can be made to recognise this by being trained on several
examples from each class. As mentioned previously, MDM templates could
be constructed from several examples to convey intraclass variance, and
with the $k$-nn variation, many templates may influence the
classification, depending on the size of $k$.

\vskip 12 pt
\centerline{\bf 4.5.\ Missing data}
\bigskip
\noindent
An important issue is how the techniques deal with missing data.  For
instance, if just a few flux measurements are missing we do not want
to have to throw away all of the data on the object (particularly if
the absences are correlated with class). In PCA, incomplete spectra
can be reconstructed through their projection onto the eigenvectors,
and more effective reconstruction techniques are also possible
(Connoly \& Szalay 1999). With MDM, the distance metric can still be
evaluated with missing dimensions, and similarly the multivariate
Gaussians of GPM can have their dimensions reduced and still provide
probabilities.  With MDM, inputs being absent is equivalent to inputs
which were never the present in the first place, because the different
dimensions of the feature vector operate independently in determining
the class. The situation is different with neural networks, because
during training all of the inputs affect the determination of all of
the weights. Thus, while a zero input will give no contribution to the
output (provided the transfer functions are symmetric about zero),
this itself may mean something, depending on what value that input had
during training.  For example, a zero input may mean a saturated
absorption line. Hence, neural networks are not particularly robust to
missing data, and the input vector should be ``completed'' in some way
before being fed to the network.

\vskip 12 pt
\centerline{\bf 4.6.\ Interpretability}
\bigskip
\noindent
A final issue in comparing these models is their interpretability. MDM
(and $k$-nn) must ``win'' as being the most obvious, although for the
probabilistically minded GPM is at least as good. I would argue that
neural networks are not nearly as obscure as people who have no
experience with them often suppose: Far from being an elaborate
``black box'', it is simply an example of a nonlinear regression
algorithm. See section 7 of Bailer-Jones et al.\ (these proceedings).

\vskip 30 pt

\centerline{\bf 5.\ A survey of the recent literature}
\bigskip
\noindent
The following survey of the literature is not intended to examine
every publication in this area. Instead it aims to illustrate how the
above methods have been used and demonstrate the performance which has
been achieved. My focus is primarily on recent work, specifically that
done in the past five years. Although earlier work played a key role
in the development and understanding of the techniques, and of
automated classification itself, this work and their results have, to
their credit, been superseded by more recent work.

I divide my survey into two main sections, MK classification and
physical parametrization.  This division is partly for convenience,
but also reflects a broad difference in the approaches initially
adopted by two communities. On the one hand, those interested in
automating MK classification have tended to use neural networks.  On
the other hand, people more concerned with stellar atmospheres and
physical parametrization have tended to use MDM. Moreover, the MK
people tend to use the entire spectrum, whereas the stellar
atmospheres people often (but not exclusively) make use of certain
line ratios and equivalent widths which are believed to have enhanced
sensitivity to physical conditions.  Furthermore, members of the
latter community have not stressed the ``automated'' side of their
work, presumably as this was not their main concern.  The emphasis on
automation from the MK community probably reflects the need to
computerize the traditional MK classification method of comparing
spectra with standards by eye.  Techniques which use the overall
appearance of a spectrum will be more robust to changes in resolution
and SNR than ones which use certain equivalent widths, because below
some resolution, equivalent widths are no longer measurable and
classification becomes impossible.  A method using ``raw'' spectral
information, on the other hand, will get less confident (higher random
errors) as the SNR or resolution degrade, but should still be able to
produce a classification. Similarly, a whole spectrum approach will
generally be more robust to missing data.

MK classification and physical parametrization are complementary
approaches.  MK classifications are fixed, while physical
parametrizations will evolve as stellar models improve. MK gives a
compact, and hence necessarily approximate, description of a stellar
spectrum.  However, as the goal is to determine the physical
parameters of stars, some kind of physical calibration is ultimately
necessary. With automated methods it is now completely feasible to
redetermine physical parameters directly from the {\it original} data
every time a new physical model is introduced, even for large
surveys. I see no problem with this approach: on the contrary, we
should always be prepared to improve our knowledge of objects as our
physical understanding grows, rather being restrained by a static
classification system.  After all, there can be no better standard
system than the underlying physics!

The performance of automated classification models is invariably
assessed from some error measure based on the residuals (the
differences between the model classifications and the ``true''
classifications), using some evaluation data set. (The ability of the
model to {\it generalize} what it has learned cannot be assessed using
the training data.)  A number of different error measures are used in
the literature, so results cannot always be easily compared.  The
RMS value of the residuals, \sigrms, is widely used. However, it is a
conservative measure in that it is dominated by the outliers, and not
necessarily representative of the majority of the residuals.  The mean
absolute value of the residuals, \mae, is more robust: because it uses only
the first power of the residuals, it is usually smaller than
\sigrms. For a Gaussian distribution,
$1\sigma$ = 1.25\mae.  Another error measure which gives a more
appropriate representation of the distribution is \sig68, the value of
the residuals which contains the central 68\% of the residuals.  It is
motivated by the fact that it is equal to the standard deviation
($\sigma$) of a Gaussian distribution (which in turn is equal to the
RMS for Gaussian distributed data). All of these errors are, of
course, just a summary of the results for all classes, and obscure any
variation of the error with class. For example,
\logg\ is typically harder to determine for cool stars than for hot stars.

Note that authors rarely state the SNR they have used, although it is
invariably high ($>$100).  Only a few articles have analysed how
performance varies with SNR. This is an important assessment for
magnitude limited surveys, because these will have the majority of
their objects at the lowest SNR.

\vskip 12 pt
\centerline{\bf 5.1.\ MK classification}
\bigskip
\noindent
PCA received early attention as a component of an automated
classifier.  Whitney (1983) used PCA to reduce the 47 spectral bins of
53 A and F stars measured over 3500--4000\AA\ to just the three most
significant principal components. A nonlinear fit to these enabled him
to determine spectral types to within 1.6 SpT (RMS error). This was no
worse than a fit using all 47 principal components, indicating how
much redundant information was present in these spectra. In comparing
this to other data compressions with PCA, one should realise that A
and F stars over this relatively narrow spectral range will show much
less variation than, say, O--M stars over a wider spectral range, so
fewer PCs will be required for a good reconstruction in the former
case.

Weaver \& Torres-Dodgen (1997) used a neural network to classify
spectra simultaneously in terms of spectral type and luminosity class,
for a range of spectral types (O--M) and luminosity classes
(I--V). This was based on high SNR 15\AA\ resolution spectra in the
range 5800--8900\AA\ which the authors had previously used for A star
classification (Torres-Dodgen \& Weaver 1995).  They used a
hierarchical system of networks: A single network first does coarse
spectral type classification. Depending on the outcome, the spectrum
is then passed to one of several more specialist networks, each of
which only knows about (i.e.\ was trained on) a subset of classes,
e.g.\ just A stars or just F stars. With such a system, the mean
absolute errors, \mae, were 0.56 SpT and 0.27 LCs (varying with
spectral type between 0.4 and 0.8 SpT, and 0.2 and 0.4 LC).  This
compares to 1.26 SpT and 0.38 LC for the coarse classifier alone, so
is quite a significant increase in accuracy for spectral type.  It
occurs because each specialist network is faced with a simpler problem
than the coarse network. It would be interesting to test whether a
single, more complex, network could achieve a performance similar to a
hierarchical approach.  Note that the system is only hierarchical in
spectral type, presumably explaining why the luminosity class accuracy
improved less. Making the structure hierarchical in all parameters
would involve a lot of networks, each of which could only be trained
on a fraction of the training data.  As training data is always
limited, there is a limit to how specialized the network structure can
be, because each network requires sufficient data to ensure that the
network weights are not underdetermined (see Bailer-Jones et al.,
these proceedings).

Bailer-Jones, Irwin \& von Hippel (1998a) used a neural network to
classify O--M spectra of luminosity classes III, IV and V, using 3\AA\
resolution spectra in the range 3500--5200\AA. Over 5000 spectra were
used, half for training and half for testing. They used PCA to
compress these 820 dimension spectra down to 25 network inputs, and
demonstrated that this compression removed noise. The spectral type
and luminosity class problems were solved separately.  A committee of
ten 25:5:5:1 networks for the spectral type problem was used. (A
committee is a system in which several identical networks are trained
from different initial random weights, and the classification results
are averaged.) The mean classification error was \sig68 = 0.82 SpT
(ranging between 0.3 and 1.0 depending on spectral type) and \sigrms =
1.09 SpT.

For the luminosity class problem Bailer-Jones et al.\ (1998a) used a
committee of ten 25:5:5:3 networks in probabilistic mode, in which
each output represents the probability that the spectrum is a member
of each class.  This achieved correct classifications for 93\% of
class III stars (giants) and 98\% for class V (dwarfs). Results for
class IV were poor (only about 10\% correct) which is worse than a
random classifier! In this case, the network is conveying the useful
information that class IV is not distinct (at least in these spectra),
which is not implausible.  Equally good results were obtained on both
the luminosity class and spectral type problems using only the line
information (using the continuum removal method of Bailer-Jones, von
Hippel \& Irwin 1998b).  Interestingly, very similar performance was also
obtained for both the spectral type and luminosity class problems when
using the entire spectrum, indicating that the PCA compression by a
factor of 33 led to no loss of classification information
(Bailer-Jones 1996). MDM was also applied to the complete spectra for
the spectral type problem, with templates formed from the average of
many training examples. Although the results were poorer (\sigrms =
2.03 SpT), only $\chi^2$ weighting was used, and class interpolation
was not (introducing a discretization error of up to 0.5 SpT), so a
direct comparison is not fair (Bailer-Jones 1996).

Singh, Gulati \& Gupta (1998) used neural networks with PCA
compression to determine spectral types of O--M stars from 11\AA\
resolution spectra in the range 3500--6800\AA.  A number of networks
with different numbers of hidden nodes and PCA inputs were tested. The
best was given by a 20:64:64:55 network used in probabilistic mode,
trained on 55 library spectra, and produced a classification error on
158 test spectra of \sigrms = 2.2 SpT.  The article by Singh et al.\
(these proceedings) shows details of the PCA compression of these
data, and compares it with the data from Bailer-Jones et al.\ (1998a).

Christlieb et al.\ (1998) used GPM to classify A5--K0 stars into one
of eight classes. The feature vector was a set of 10 line strengths
measured from optical spectra from the Hamburg/ESO objective prism
survey. The model was trained using the EM (expectation maximization)
algorithm on 671 spectra and tested on the same data using the
leave-one-out-method (i.e.\ 670 separate models are trained on each
combination of 670 spectra, and the performance evaluated on the one
left out). The overall misclassification rate was 28\%, but only 1\%
of objects were incorrectly classified by more than one class.

The above examples concern classification in visual blue and
red spectra, but work has also shown that MK classification is
possible in the ultraviolet.  Vieira \& Pons (1995) used
extinction-corrected IUE spectra (1150--3200\AA) at 2\AA\ sampling in
a neural network classifier, and could achieve an RMS accuracy of 1.1
SpT for spectral types O3 to G5. Almost identical results were
obtained with an unweighted MDM classifier.

\vskip 12 pt
\centerline{\bf 5.2.\ Physical parametrization}
\bigskip
\noindent
Vansevicius \& Bridzius (1994) used MDM with $\chi^2$ weighting to
determine spectral type and absolute magnitude, $M_V$, from six colour
indices defined from Vilnius photometry. RMS accuracies of 0.7 SpT
and 0.8 mag respectively for O5--M5 stars with $-9 < $~M$_V < +12$ were
obtained. They further attempted to determine colour excess
[(B$-$V)$-$(B$-$V)$_0$] from the value of the reduced $\chi^2$
(evaluated from $D_c$ in equation 1), based on the belief that this
should equal 1.0 for no reddening. While an intriguing idea, this
assumption appears to ignore the existence of intraclass variation.

Bailer-Jones et al.\ (1997) used a neural network to physically
calibrate spectra in terms of effective temperature, \teff. They
calculated a grid of synthetic spectra for a range of effective
temperatures and surface gravities, and processed them to have the
same properties (wavelength sampling and flux scale) as the observed
spectra (those used in Bailer-Jones et al.\ 1998a).  An 820:5:5:1
neural network was trained on these synthetic spectra, so that when
the real spectra were applied, effective temperatures were determined
directly.  As these real spectra had known spectral types, it was also
possible to derive an accurate \teff--SpT calibration for giants,
subgiants and dwarfs. They further showed that the calibration was
metallicity dependent, indicating \met~$= -0.2$ for the sample.
Gulati, Gupta \& Rao (1997) independently used the same approach to
calibrate G and K dwarf stars from 4850--5380\AA\ spectra at a
resolution of 2.4\AA. Their spectra had already been assigned \teff\
by other means, and they showed that the network could reproduce these
to $\pm 250$~K. This is an upper limit, set by the sampling of \teff\
in the synthetic spectral grid. An MDM method with no weighting was
similarly grid limited.

Katz et al.\ (1998) used MDM with $\chi^2$ weighting to determine
\teff, \logg\ and [Fe/H] of high resolution (0.1\AA) echelle
spectra over the range 3900--6800\AA. The templates were synthetic
spectra calculated with temperatures between 4000~K and 6300~K, \logg\
between 0.6 and 4.7 dex and metallicities between $-2.9$ and $+0.35$
dex. With a SNR of 100, the RMS errors were log~\teff = 0.008, \logg\
= 0.28, and [Fe/H] = 0.16.  If the SNR was degraded to 10, the errors
were no worse (log~\teff\ = 0.009, \logg\ = 0.29, and [Fe/H] = 0.17).
(For reference, an error of $\sigma$ in log co-ordinates is a
fractional error of $2.3\sigma$\% in linear co-ordinates.)

Bailer-Jones (2000) used a neural network approach to determine
all three principal stellar parameters over a wide parameter
range (\teff\ = 4000--30000~K,
\logg\ = 2.0--5.0, \met\ = $-3.0$ to $+1.0$) working only
with synthetic spectra (3000--10000\AA). The 3000 synthetic spectra
were randomly split into two sets, the network trained on one and its
performance tested on the other. The goal was to investigate the
effect of SNR and resolution on the ability to
determine the parameters. The surprising result was that even at low
resolution (FWHM of 50--100\AA) and SNR (5--10 per resolution element),
\teff\ and \met\ could be determined to 1\% and 0.2 dex respectively, 
and \logg\ to 0.2 dex for stars earlier than solar (mean absolute
errors), using a 35:5:10:3 network.  The spectra retained absolute
flux information, because the simulations were done for the GAIA
parallax mission. This certainly helps the \teff\ determination (as $L
= 4 \pi R^2 \sigma $\teff$^4$), but the network still has to disentangle
the influences of the three parameters on the spectra. The work also
tested a number of proposed GAIA filter systems.

Again in the context of simulating performance for a parallax mission
(this time DIVA), Elsner et al.\ (1999) used an unweighted MDM method
to determine \teff\ and \logg\ from low resolution synthetic spectra
(varying from 160--380\AA\ FWHM over the range 3130--9990\AA).
Simulating objects at $V=12$, they obtained RMS errors of 0.5 dex in
\logg\ and 10\% in \teff\ for hot stars, and 0.15 dex and 5\%
respectively for cool stars. With the assumed design for DIVA, this
precision (or better) would apply to about 2.5 million stars.

Snider et al.\ (2001) applied neural networks to the physical
parametrization of 264 observed spectra with \teff $< 6500$~K.  These
were observed at 2\AA\ resolution over the range 3630--4890\AA.  The
data were initially calibrated by physical methods, and a subset used
for training a network with 1952:5:3 architecture.  The RMS errors
were 3\% in \teff, 0.27 dex in \logg\ and 0.22 dex in
\met. These results are comparable to those
obtained by Katz et al., but using a smaller wavelength coverage and
much lower resolution.

Allende Prieto \& Lambert (2000) have developed a method for
estimating masses and ages from Hipparcos data using evolutionary
models. The Hipparcos parallax, fluxes and B$-$V colour enable a
determination of the effective temperature and radius of the star
(once a suitable bolometric correction is applied).  The position of
the star in the theoretical HR diagram is then compared to theoretical
evolutionary tracks for objects of different masses. By averaging over
all possible tracks which lie within the error box of the object, an
estimate of the mass and age (with uncertainties) is obtained.  The
method was tested against objects with known parameters from eclipsing
spectroscopic binaries, with RMS errors of 12\% in mass, 6\% in radius
and 4\% in \teff. A better mass determination (8\% error) is possible
if the metallicity of the objects is known (assumed near-solar in this
case). When combined, these errors correspond to an error in \logg\ of
about 0.06 dex. This is considerably lower than that obtained by other
methods, and indicates that evolutionary models could be useful in
constraining the possible surface gravities. However, it should be
pointed out that the \logg\ of the sample only varied between 3.7 and
4.5 dex.

\vskip 12 pt
\centerline{\bf 5.3.\ Other issues}
\bigskip
\noindent
Interstellar extinction (reddening) will affect any deep survey, and
it is important to ensure that it does not bias determinations of
stellar parameters. One approach to circumvent this is to determine
the degree of extinction from the spectra, with the aim of correcting
for it.  Gulati, Gupta \& Singh (1997) attempted such a determination
with both a neural network and $\chi^2$ weighted MDM, using the
interstellar absorption feature at 2200\AA. From 6\AA\ resolution
spectra they were able to determine E(B$-$V) to within 0.08 magnitudes
(RMS error) over the range 0.05--0.95 magnitudes. At least some of
this error is the discretization error in the template/training
spectra, which had E(B$-$V) in steps of 0.05 magnitudes.

Another important issue related to classification is the fact that
many stellar systems are binaries, and -- as most will not be
resolvable -- will have composite spectra.  Whenever the brightness
ratio is not extreme, it is in principle possible to determine the
parameters of both components. Weaver (2000) has attempted to do this
with a neural network which has two sets of outputs, one for each
component. The network had an additional output for determining the
brightness ratio.  Composite spectra were artificially constructed by
combining the spectra of Weaver \& Torres-Dodgen (1997).  Mean
absolute classification errors for both components of 2.5 SpT and 0.45
LC could be obtained, although this average masks a large dependence
on brightness ratio.  As the brightness ratio increased from 1 to 20,
the errors varied from about 1.7 to 7 SpT, and 0.2 to 1.0 LC.  Some
improvement was obtained with a hierarchical system (i.e.\ using
networks specialized in classifying, say, only A--F, or G--K
binaries). The network included recurrent feedbacks from the output
layer to the input, with the justification that this improved the
results, but it was not clear why this should be the case.  Note that
the network assumed all spectra to be binaries, so assigns two classes
to all spectra, even if it is not composite. Clearly, with this
approach, some kind of preprocessor is necessary to determine which
spectra really are binary.

\vskip 12 pt
\centerline{\bf 6.\ Summary of the current status of automated classification}
\bigskip
\noindent
I will now summarise where we are in terms of our ability to classify
stars automatically in line with the goal given in section 2.

\noindent 1. All three physical parameters (\teff, \logg, \met) can be
determined to reasonable accuracy from spectra in an automated fashion.

\noindent 2. We need neither high resolution spectra nor high
signal-to-noise data.  A resolution of around 100\AA\ FWHM and SNR of
about 10 per resolution element appear to be sufficient, although a
relatively large wavelength coverage (several thousand \AA)\ may then
be necessary. Such a wide coverage may well be desirable for many
surveys anyway, and presents no particular technical problem if
obtained photometrically.

\noindent 3. Relatively simple classification models are sufficient, for example
neural networks with one or two hidden layers each with 5--10 hidden
nodes or MDM with simple weighting schemes.

\noindent 4. For a wide range of stars (O--M, I--V), spectral types
can be determined to 0.3--0.8 SpT and luminosity classes to 0.2--0.4
LC (or 95\% correct classification rate).

\noindent 5. Physical parameters can be determined directly from spectra by 
training neural networks on synthetic spectra (for
example). \teff\ and \met\ can be obtained with relative ease to 1\%
and 0.2 dex respectively across a wide parameter range. \logg\ can be
determined to about 0.2 dex for early type stars, but only 0.5 dex for
later-type stars.

\noindent 6. Parallax information is very useful, as it enables a determination 
of absolute luminosity and hence radius. Thus it is particularly
important that powerful automated stellar classification systems are
developed for parallax missions such as DIVA, FAME and GAIA.

\noindent 7. We can make some attempt to estimate the fundamental parameters
of age and mass through the combination of evolutionary models and
classification models.

\noindent 8. Some object identification (discrimination of stars from other objects),
binary classification and interstellar extinction determination is
possible.

\vskip 30 pt

\centerline{\bf 7.\ Future requirements}
\bigskip
\noindent
I hope to have shown that many of the constituents of a survey
classification system have been demonstrated. However, much work
remains to be done to develop and implement a complete system, and I
finish this review by highlighting what further developments are
required.

To date, all applications have used ``cleaned'' data sets of
preselected objects.  This will not be the case for a blind survey,
and a robust system for identifying which objects are stellar is
essential. Methods which have been demonstrated include neural
networks for star--galaxy separation based on image data (e.g.\
Odewahn et al.\ 1993) and the recognition of non-stellar objects from
the projection of spectra onto stellar principal components
(Bailer-Jones et al.\ 1998a).

Stellar spectra show much more variation than is represented by the
three parameters \teff, \logg\ and \met. Thus classifiers need an
extended parameter space to include factors such as microturbulence and
the $\alpha$ abundance ratios. Similarly,
different phases of evolution, such as pre-main-sequence stars and
white dwarfs, as well as peculiar stars and pulsating stars, need to
be catered for, as do extensions to ultra cool objects (L and
T dwarfs). Developments in stellar structure and atmospheric models
are required to model real spectra better, for example to take account
of non-LTE effects, chromospheres and the formation of dust at low
temperatures. Many such developments are in progress or could already
be incorporated.

Large data sets will inevitably have missing data, and classifiers
must be able to cope with this in a robust way. Some theoretical
discussion was given in section~4, but there has been little empirical
evaluation of this problem. Additionally, the classification system
should be able to assess its own uncertainties when making
classifications (whether inputs are complete or not), and not just
rely on global statistical estimates from test data. For example, with
a neural network, uncertainties will generally be higher where the
training data are sparse, and methods exist for quantifying these
(e.g.\ MacKay 1995). Local error measures appear to be easily
obtainable in both the MDM and GPM methods.

Finally, I mention the need for the automated classification system to
be designed in parallel with the survey.  Just as the classification
system must be able to cope with what can be measured, and not make
unrealistic demands, the survey must provide the data which are
required by the classification system for producing reliable
classifications. The products of large surveys will be of a more
statistical nature than has previously been the case, so greater
interplay between classifier and survey development is essential.

This review has focused on just four methods, yet they are essentially
the only methods which have been used in automated stellar
classification. Each method has its own advantages and disadvantages,
and it is not my intention to point to a ``winner'', particularly as
GPM has not seen much application to stellar classification, further
optimization is desirable with MDM, and entirely distinct methods
exist which have not even be tested.  Indeed, the exploration of fully
alternative approaches is of much interest.  But generally speaking,
for multidimensional problems, it appears that some kind of
interpolation method is appropriate, rather than a look-up table type
approach. Also, as both the variety of spectra encountered and the
number of parameters we wish to determine increase, a limited
hierarchical approach may be useful.

I would like to acknowledge the following people for useful
conversations at various times over the past few years: Norbert
Christlieb, Chris Corbally, Bob Garrison, Richard Gray, Gerry Gilmore,
Mike Irwin, Ofer Lahav, Michael Storrie-Lombardi and Ted von Hippel.

\bigskip
\centerline{\bf References}
\bigskip
{\eightpoint\parindent=0pt\everypar={\hangindent=0.5 cm}

Allende Prieto, C., Lambert, D.L., 1999, A\&A, 352, 555

Bailer-Jones, C.A.L., 1996, PhD thesis, Univ.\ Cambridge

Bailer-Jones, C.A.L., 2000, A\&A, 357, 197

Bailer-Jones, C.A.L., Irwin, M., Gilmore, G., von Hippel T., 1997, MNRAS, 292, 157

Bailer-Jones, C.A.L., Irwin, M., von Hippel T., 1998a, MNRAS, 298, 361

Bailer-Jones, C.A.L., Irwin, M., von Hippel T., 1998b, MNRAS, 298, 1061

Christlieb, N., Grasshoff, G., Nelk, A., Schlemminger, A., Wisotzki, L., 1998,
in Classification, Data Analysis and Data Highways, I. Balderjahn et al.\ (eds.),
Springer: Berlin, p.\ 16

Clowes, R., Adamson, A., Bromage, G., 2001, The New Era of Wide-Field
Astronomy, ASP Conf.\ Ser., Astronomical Society of the Pacific: San
Francisco, to appear

Connolly, A.J., Szalay A.S., 1999, ApJ, 117, 2052

Elsner, B., Bastian U., Liubertas, R., Scholz, R., Baltic Astronomy, 8, 385

Goebel, J., Volk, K., Walker, H., Gerbault, F., Cheeseman, P., Self,
M., Stutz, J., Taylor, W., 1989, A\&A, 222, L5

Gulati, R.K., Gupta R., Rao, N.K., 1997, A\&A, 322, 933

Gulati, R.K., Gupta R., Singh, H.P., PASP 109, 843

Katz, D., Soubiran, C., Cayrel, R., Adda, M., Cautain, R., 1998, A\&A, 338, 151

MacKay, D.J.C., 1995, Network: Computation in Neural Systems, 6, 469

Morgan, W.W., Keenan, P.C., Kellman, E., 1943, An Atlas of Stellar
Spectra with an Outline of Spectral Classification, University of
Chicago Press: Chicago

Odewahn, S.C., Humphreys, R.M., Aldering, G., Thurmes, P., 1993, PASP, 105, 1354

Singh, H.P., Gulati, R.K., Gupta, R., 1998, MNRAS, 295, 312

Snider, S., Qu, Y., Allende Prieto, C., von Hippel, T., Beers, T.C.,
Sneden, C., Lambert, D.L., Rossi, S., 2001, ASP Conf.\ Ser., 223,
CD-1344

Vansevicius, V., Bridzius, A., 1994, Baltic Astronomy, 3, 193

Vieira, E.F., Pons O.D., 1995, A\&AS, 111, 393

Weaver, W.B., 2000, ApJ, 541, 298

Weaver, W.B., Torres-Dodgen, A.V., 1995, ApJ, 446. 300

Weaver, W.B., Torres-Dodgen, A.V., 1997, ApJ, 487, 312

Whitney, C.A., 19832, A\&AS, 51, 443

}
\end